\documentclass[twocolumn]{aastex701}
\usepackage{amsmath}

\DeclareUnicodeCharacter{2212}{-}

\begin{document}

\title{ The Far-Ultraviolet Extragalactic Legacy (FUEL) Survey:

Hubble Far-UV Images and Catalogs of the Extragalactic Legacy Fields}

\author[0009-0003-4111-6472]{Aliakbar Kavei}
\affiliation{Department of Physics and Astronomy, University of California, Riverside, CA 92521, USA}
\email[show]{aliakbar.kavei@email.ucr.edu}  

\author[0000-0002-4935-9511]{Brian Siana}
\affiliation{Department of Physics and Astronomy, University of California, Riverside, CA 92521, USA}
\email{brian.siana@ucr.edu}

\author[0000-0002-7064-5424]{Harry Teplitz}
\affiliation{IPAC, California Institute of Technology, 1200 E. California Blvd, Pasadena, CA 91125, USA}
\email{hit@ipac.caltech.edu}

\author[0000-0002-8630-6435]{Anahita Alavi}
\affiliation{IPAC, California Institute of Technology, 1200 E. California Blvd, Pasadena, CA 91125, USA}
\email{anahita@ipac.caltech.edu}

\author[0000-0002-3433-4610]{Alberto Dominguez}
\affiliation{IPARCOS and Department of EMFTEL, Universidad Complutense de Madrid, E-28040 Madrid, Spain}
\email{alberto.d@ucm.es}

\author[0000-0001-9491-7327]{Simon P. Driver}
\affiliation{International Centre for Radio Astronomy Research (ICRAR) and the International Space Centre (ISC), University of
Western Australia, Crawley, WA 6009, Australia}
\email{simon.driver@uwa.edu.au}

\author[0000-0001-8419-3062]{Alberto Saldana-Lopez}
\affiliation{Department of Astronomy, Oskar Klein Centre, Stockholm University, 106 91 Stockholm, Sweden}
\email{alberto.saldana-lopez@astro.su.se}

\author[0000-0001-6482-3020]{James Colbert}
\affiliation{IPAC, California Institute of Technology, 1200 E. California Blvd, Pasadena, CA 91125, USA}
\email{colbert@ipac.caltech.edu}

\author[0000-0001-5091-5098]{Joel Primack}
\affiliation{Department of Physics, University of California at Santa Cruz, Santa Cruz, CA 95064, USA}
\email{joel@ucsc.edu}

\author[0000-0002-6584-1703]{Marco Ajello}
\affiliation{Department of Physics and Astronomy, Clemson University, Kinard Lab of Physics, Clemson, SC 29634, USA}
\email{majello@clemson.edu}

\begin{abstract}

We present far-ultraviolet (FUV) images and catalogs from the Hubble Space Telescope (HST) Advanced Camera for Surveys/Solar Blind Channel (ACS/SBC) F150LP ($\sim 1600$ \AA) of three extragalactic fields: GOODS-S, GOODS-N, and COSMOS. 
The data comprise 365 orbits of high-resolution imaging of 151 pointings covering an area of 44.7 arcmin$^2$ to typical depths of FUV $\approx 28.7$ AB ($3\sigma$, $0.5''$ diameter aperture). 
We provide a new model of the spatially-varying dark ``glow'' created from all 365 orbits of data, and scale and subtract it from all pointings. 
We provide drizzled image mosaics, weight maps, and exposure time maps matched in coordinates and pixel scale to the Hubble Legacy Fields (HLF) frame, and original COSMOS tiles. 
Galaxy photometry is measured within isophotes defined with existing deep \textit{Hubble} F606W or F814W optical filters. 
We detect 1068 galaxies and provide catalogs of all optical detections, and include matched IDs to existing 3D-HST and CANDELS catalogs. 
The redshift distribution of FUV-detected galaxies peaks at $z\sim0.6$ and falls until $z=1.2$, where the Lyman limit shifts redward of any filter transmission. 
These data fill the redshift gap of high-resolution far-UV imaging between $z\sim0$ and $z>1$, allowing for studies of star-forming regions, dust properties, the FUV extragalactic background, and Lyman continuum emission from $z>1.2$ galaxies.

\end{abstract}


\section{Introduction}

Deep, high-resolution, rest-frame far-ultraviolet (FUV) imaging directly traces the youngest stellar populations and the geometry of dust on sub-kpc scales. GALEX provided wide-area FUV maps in the local universe but with coarse resolution \citep[$4.5''$,][]{2005ApJ...619L...1M} that was not sufficient for resolving sub-kpc structure in galaxies outside of the Local Group. {\it Hubble Space Telescope} (HST) near-UV and optical imaging and {\it James Webb Space Telescope} (JWST) near-IR imaging allow sub-kpc rest-frame FUV imaging of galaxies at $1<z<6$ and $z>6$, respectively. At intermediate redshifts ($0.2<z<0.9$), HST FUV imaging provides the angular resolution needed for sub-kpc, spatially resolved studies. In this redshift range, the F150LP filter probes rest-frame $\sim850$--$1350$~\AA\ and therefore fills a key observational gap between GALEX studies in the local universe and higher-$z$ rest-frame FUV studies enabled by HST/JWST at longer observed wavelengths. However, in this important epoch spanning the second half of the universe, existing HST FUV imaging is sparse and fragmented. Prior HST FUV imaging has largely been confined to very small areas (e.g., the Hubble Deep Field North \citep{Teplitz2006} and the Hubble Ultra Deep Field \citep{2007ApJ...668...62S}), limiting our ability to study important aspects of galaxy evolution. As a result, our knowledge of the sizes, ages, and dust extinction of star-forming regions is often better constrained in the earlier half of the universe than in the latter half, motivating deep, high-resolution observed-frame far-UV imaging over a sufficiently large area to detect large numbers of galaxies.

The ACS/SBC is the wider of the two FUV imagers on Hubble, with $1.7\times$ the field of view of the Space Telescope Imaging Spectrograph (STIS) imager, and is therefore typically used for mapping larger fields (e.g., \cite{Teplitz2006, 2007ApJ...668...62S}). The F150LP long-pass filter (central wavelength 1606~\AA; see \cite{2007ApJ...668...62S}) has often been the preferred filter for general far-UV imaging surveys, as the long-pass cutoff at $\sim1500$\AA\ avoids the brightest geocoronal lines, but still has high system throughput before the detector sensitivity falls at longer wavelengths. In combination with the exceptional ancillary data in GOODS and COSMOS, these mosaics enable community studies of outstanding questions in galaxy evolution.

Uniform reduction of existing HST/ACS/SBC F150LP imaging has been impeded by two issues: (1) a detector-position- and detector-temperature-dependent dark ``glow'' that complicates background modeling, and (2) the small field of view and the low surface density of UV-bright sources, which reduces the number of high-S/N objects for reliable alignment and mosaicing. Moreover, existing GO programs (individual HST observing programs led by different teams) have adopted different reduction choices (e.g., dark modeling/background subtraction, drizzling parameters, and astrometric registration), which makes the resulting products heterogeneous and difficult to combine into a consistent mosaic and catalog. These factors have historically limited the availability of uniform public FUV products beyond a few deep pointings (e.g., HDF-N and the UDF).

Despite these challenges, the archive already contains a de facto survey: over the past $\sim$18 years, seven programs have obtained 365 orbits in 151 ACS/SBC pointings across the GOODS-North, GOODS-South \citep{2004ApJ...600L..93G}, and COSMOS \citep{Capak_2007} fields. Uniform mosaics have been obtained and are available in the archive from specific programs covering the Hubble Deep Field North \citep{Teplitz2006} and the Hubble Ultra Deep Field \citep{2007ApJ...668...62S}. However, only about 21\% of the observations have previously been uniformly reprocessed into science-ready, co-registered mosaics and released, while many targeted and parallel observations remain only as standard pipeline-calibrated exposures and have not been uniformly background/dark-corrected, aligned, or combined into mosaics. Because these existing FUV images cover a large area ($\sim45$ arcmin$^2$) and are concentrated in legacy extragalactic fields with superb ancillary data, these data provide an opportunity to fill the intermediate-$z$ FUV gap. This effort is complementary to Hubble archival programs that provide uniformly reduced mosaics and catalogs in the optical and near-infrared (e.g., the Hubble Legacy Fields; \citealt{2016arXiv160600841I, Whitaker_2019}).
However, these products do not include a similarly uniform set of far-UV mosaics and source catalogs, which we present here.

This compilation therefore enables substantially improved constraints on the far-UV extragalactic background light (EBL) from integrated galaxy light. Current estimates of faint far-UV counts that comprise a large fraction of the integrated light are based on ACS/SBC imaging over an extremely small area in only two fields \citep[e.g.,][]{Voyer_2011} and are therefore subject to significant Poisson uncertainty and cosmic variance. By adding a third independent field (COSMOS) and increasing the effective area to $\sim45$~arcmin$^2$ (2.8$\times$ larger than previous {\it Hubble}-based counts), the resulting catalogs will considerably reduce Poisson uncertainty and mitigate cosmic variance. 

Recent direct measurements of the cosmic {\it optical} background with {\it New Horizons} underscore the need for accurate, deep number counts to interpret absolute background levels. Early analyses reported an excess above the integrated light from cataloged galaxies (e.g., \citealt{2018Sci...362.1031F, Lauer_2022}), but subsequent work suggests that improved treatment of faint-galaxy contributions and observational systematics can largely reconcile the measurement with galaxy counts (e.g., \citealt{2024ApJ...972...95P}). Although these results are at optical wavelengths, the same principle applies in the UV: well-characterized faint-galaxy counts are critical for an accurate determination of the integrated (resolved) contribution to the short-wavelength EBL and for constraining any additional diffuse or unresolved emission.

In addition, the deep, high-resolution ACS/SBC images are ideal for LyC searches at $1.2<z<1.5$. Previous SBC LyC work at these redshifts included targeted observations of 26 galaxies \citep[e.g.,][]{Siana_2010, Alavi_2020} as well as 21 lower-mass galaxies in the HDF-N and HUDF (e.g., \citealt{2007ApJ...668...62S}). Today, these legacy fields contain thousands of spectroscopic redshifts from major surveys, enabling LyC measurements of much larger samples and stacked constraints on average escape fractions.

Here we aggregate all ACS/SBC F150LP data in the greater GOODS-N, GOODS-S, and COSMOS fields; register every exposure to the common WCS of existing optical/near-IR high-level products; construct a model of the temperature-dependent dark glow with per-exposure scaling; produce uniformly processed mosaics; and create FUV catalogs. This includes the data from previously released images in the HDF-N \citep{Teplitz2006} and the HUDF \citep{2007ApJ...668...62S}, but applying the latest calibrations and using an improved dark current model. We adopt a recent ACS/SBC zeropoint update (see Sections~\ref{sec: Source Detection and Photometry} and \ref{sec: Results} for details). The resulting images and catalogs are released via the Mikulski Archive for Space Telescopes (MAST) as a High-Level Science Product (HLSP; \dataset[FUEL-SURVEY]{\doi{10.17909/2yqw-3g14}}).

These products enable a variety of science investigations, including constraints on the FUV extragalactic background light (e.g., \citealt{Voyer_2011}); measuring the Lyman-continuum escape fraction at $1.2<z<1.5$ (e.g., \citealt{2007ApJ...668...62S, Siana_2010, 2010ApJ...720..465B, Alavi_2020}); and resolved studies of dust and clumpy/bursty star formation at $0.2<z<0.9$ (e.g., \citealt{Teplitz2006, 2019MNRAS.484.4897C, Martin_2023}), including the sizes and ages of star-forming clumps and their evolution, resolved dust maps to better understand dust attenuation curves, and timescales of ``bursty'' star formation (e.g., \citealt{2015MNRAS.451..839D}). More generally, these far-UV products add significant value to funded legacy efforts that compile the near-UV, optical, and near-IR imaging in these fields (e.g., \citealt{2021MNRAS.507.5144S,2022AJ....164..141W,2025arXiv250703412T}), completing the wavelength range accessible to HST.

This paper is organized as follows. Section~\ref{sec:Observations and Data Overview} summarizes the existing archival data set. Section~\ref{sec:Data Reduction and Image Processing} describes the reduction and alignment, including dark-glow modeling and variance weighting. Section~\ref{sec: Source Detection and Photometry} details source detection and photometry. Section~\ref{sec:catalog-construction-and-description} presents the details of the catalog. Section~\ref{sec: Results} reports depth and quality assessments. Section~\ref{sec:Applications} outlines key scientific applications enabled by these data. Section~\ref{sec:summary} summarizes the release and provides access instructions.

\section{Observations and Data Overview} \label{sec:Observations and Data Overview}

\begin{figure*}[htb!]
\centering
\includegraphics[width=1\textwidth]{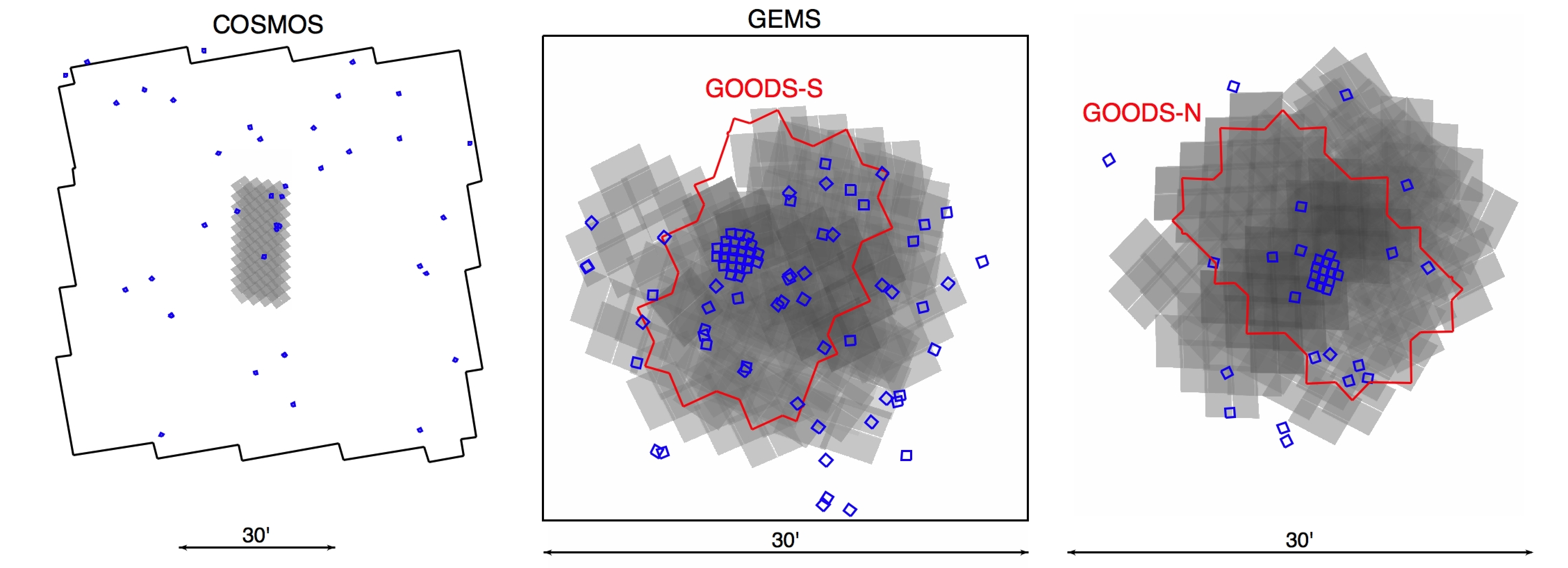}
\caption{Footprints of SBC/F150LP observations in the COSMOS, GOODS-N, and GOODS-S fields, overlaid on CANDELS \citep{2011ApJS..197...35G, 2011ApJS..197...36K} ACS/WFC F814W exposure maps (gray scale). Blue squares mark ACS/SBC far-UV pointings, red outlines indicate the footprint of the ACS optical data from the GOODS Survey \citep{2004ApJ...600L..93G}, and black outlines show wide-field optical HST/ACS imaging in COSMOS and GEMS \citep{2004ApJS..152..163R}. The scale bar in each panel indicates 30 arcminutes.}
\label{fig:SBC_FOV}
\end{figure*}

We analyze archival ACS/SBC FUV (F150LP) imaging obtained through seven HST programs covering COSMOS, GOODS-N, and GOODS-S. The spatial footprints of these observations are presented in Figure~\ref{fig:SBC_FOV}, with the program details summarized in Table~\ref{tab:programs}.

\begin{table*}[htb!]
\centering
\caption{List of programs with ACS/SBC far-UV (F150LP, 1606~\AA) imaging in the COSMOS, GOODS-N, and GOODS-S fields.}
\label{tab:programs}
\renewcommand{\arraystretch}{1.2}
\setlength{\tabcolsep}{10pt}
\begin{tabular*}{\textwidth}{@{\extracolsep{\fill}} l c c c l l }
\hline
Program ID & PI & Pointings & Orbits & Proposed Science & Data Products \\
\hline
9478            & Teplitz    & 14  & 28  & HDF-N Legacy & Images/Catalogs \\
10403           & Teplitz    & 25  & 50  & UDF Legacy        & Images \\
10403 (parallel)& Teplitz    & 2   & 12  & Parallel          & None \\
10872           & Teplitz    & 15  & 75  & $f_{\mathrm{esc}}$ at $z\sim1.3$ & None \\
11082           & Conselice  & 29  & 81  & Parallel          & None \\
11144           & Bouwens    & 23  & 61  & Parallel          & None \\
11236           & Teplitz    & 32  & 32  & $f_{\mathrm{esc}}$ at $z\sim0.7$ & None \\
14123           & Colbert    & 11  & 26  & $f_{\mathrm{esc}}$ at $z\sim1.3$ & None \\
\hline
Data Products    &            & 39  & 78  &                   & \\
No Data Products &            & 112 & 287 &                   & \\
\hline
Total            &            & 151 & 365 &                   & \\
\hline
\end{tabular*}
\end{table*}

\begin{figure}[htb!]
\centering
\includegraphics[width=0.45\textwidth]{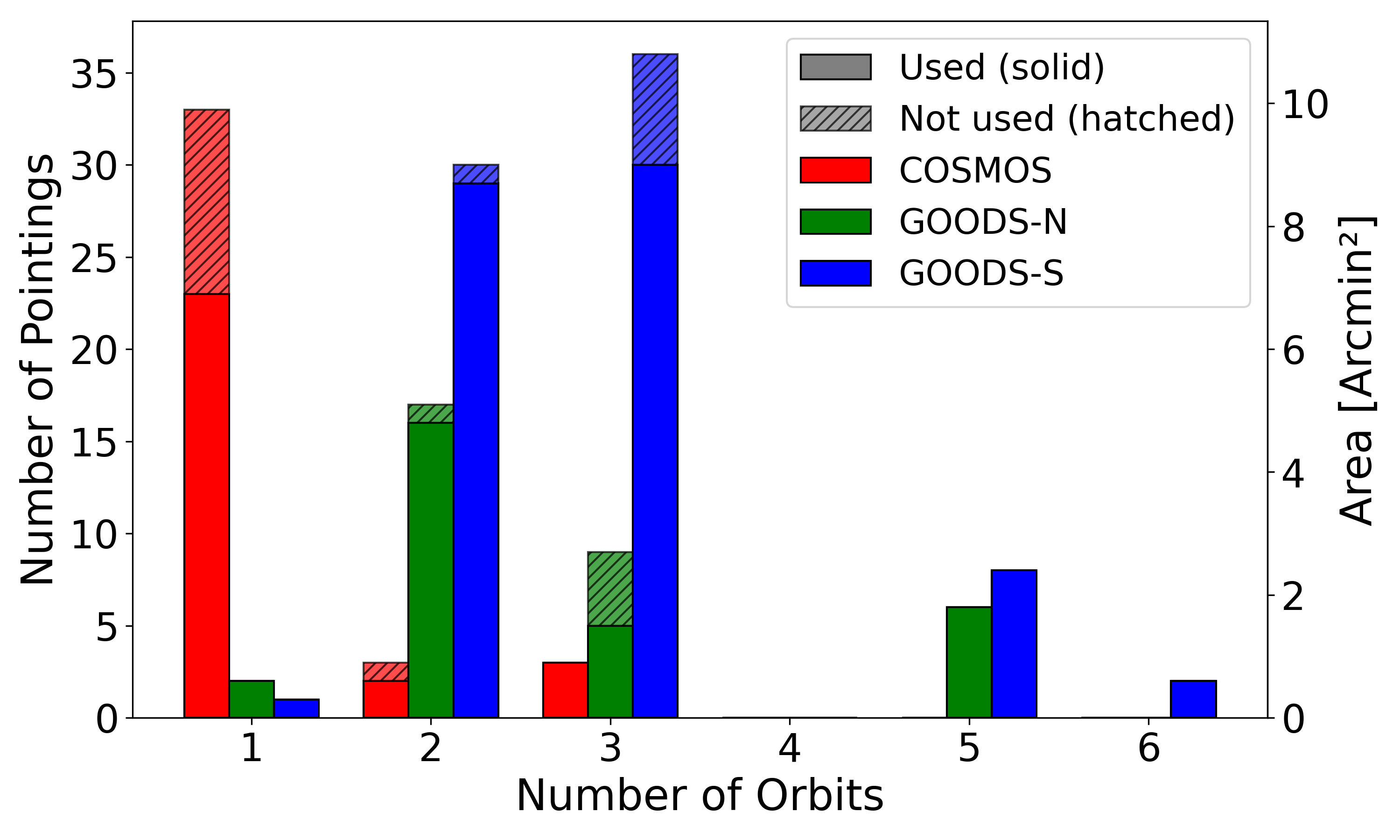}
\caption{Histogram showing the number of ACS/SBC pointings (left axis) and cumulative area (right axis) as a function of exposure time in orbits for COSMOS (red), GOODS-N (green), and GOODS-S (blue). Most pointings are 2-3 orbits in depth, with 16 pointings reaching depths of 5-6 orbits.}
\label{fig:depthhist}
\end{figure}

The cumulative area covered by these observations is approximately 45 square arcminutes. Typical exposure times range from 2--3 orbits per pointing (roughly $\sim$5{,}000--8{,}000 s), with a subset of 16 deeper integrations of 5--6 orbits (roughly $\sim$13{,}000--16{,}000 s). Figure~\ref{fig:depthhist} illustrates the distribution of exposure times across the surveyed fields. About half (77 of 151) of the pointings are in the GOODS-S Field, with similar numbers of pointings in GOODS-N (34) and COSMOS (40). 33 of the 36 shallow (1-orbit) pointings are in the COSMOS field, from GO-\texttt{11236} (\citealt{2010ApJ...720..465B}).

\section{Data Reduction and Image Processing} \label{sec:Data Reduction and Image Processing}

The reduction of the SBC data requires a number of additional steps beyond what is typical when reducing {\it Hubble} optical/near-IR imaging, including special dark ``glow'' subtraction, and alignment by hand to the few high S/N detections in any given image. The steps are similar to those of \citet{Teplitz2006} and \citet{2007ApJ...668...62S} in the HDF-North and HUDF to produce those high-level science products.

\subsection{Alignment} \label{subsec:Data Reduction and Image Processing}

To accurately align the ACS/SBC images with corresponding optical \citep[F606W and F814W, ][]{Whitaker_2019, Koekemoer_2007, 10.1111/j.1365-2966.2009.15638.x} and near-UV \citep[F275W and F336W, ][]{Teplitz_2013, Rafelski2015UVUDF} images, we specifically aligned the GOODS fields to the mosaics provided by the Hubble Legacy Fields collaboration\footnote{\url{https://archive.stsci.edu/prepds/h f/}} \citep{2016arXiv160600841I}, as these data include all of the pointings outside of the formal GOODS footprints, where many of the ACS/SBC pointings lie. For the COSMOS field, alignment was performed using the publicly available COSMOS datasets\footnote{\url{https://irsa.ipac.caltech.edu/data/COSMOS/images/acs_mosaic_2.0/}}. The data consist of 151 pointings, each consisting of multiple dithered exposures. We first combine the individual exposures of each SBC pointing using the \textit{Drizzle} technique \citep{2002PASP..114..144F, 2012drzp.book.....G}, adopting the default WCS solution in the pipeline-calibrated SBC exposure headers as the initial astrometric reference. The input exposures to this drizzle process are already flat-fielded (we start from the archive calibrated \texttt{\_flt} products produced by the standard \texttt{CALACS} pipeline; the relevant pipeline/CRDS versions are recorded in the FITS header keywords \texttt{CAL\_VER} and \texttt{CRDS\_VER}), but have not been dark-subtracted.

\begin{figure}[htb!]\centering\includegraphics[width=0.46\textwidth]{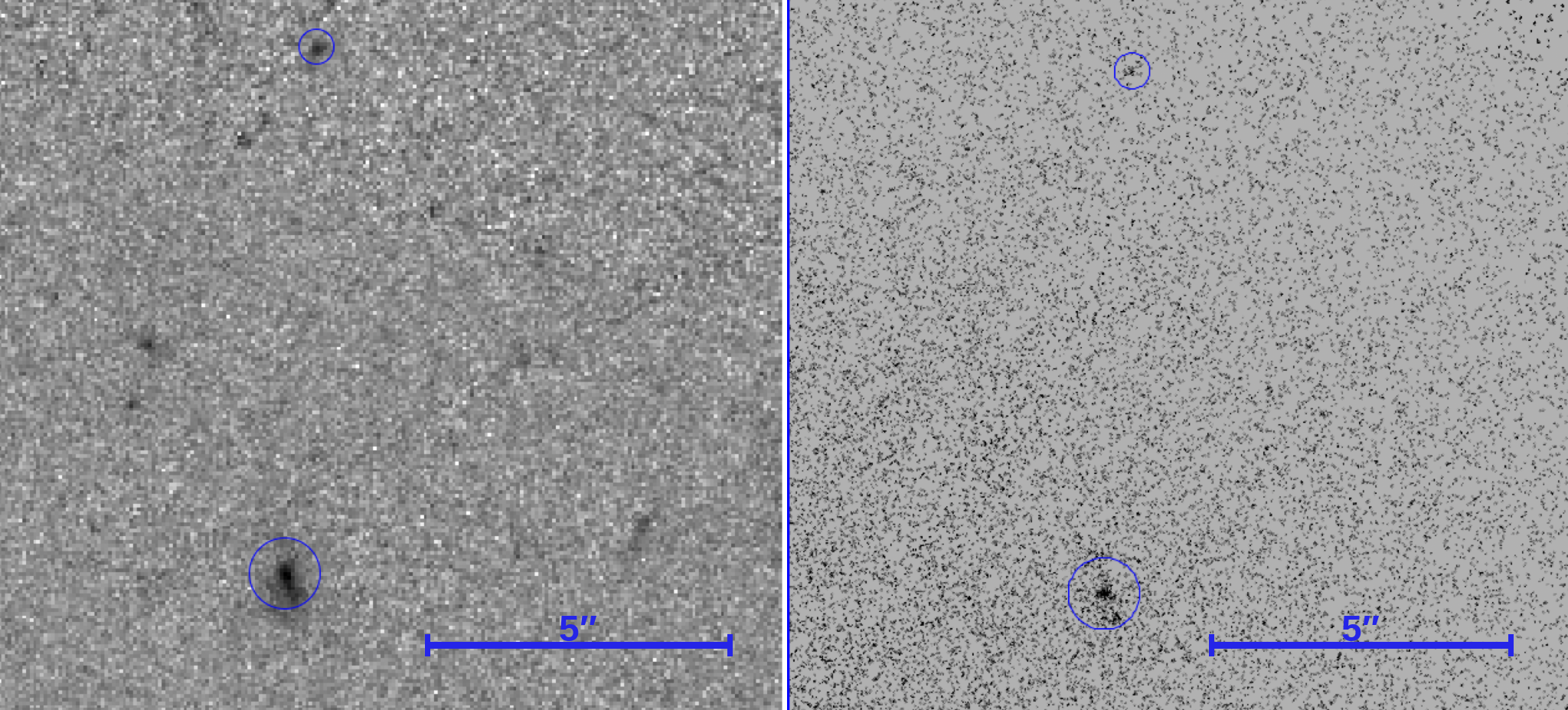}\caption{A small region in GOODS-South imaged for two orbits with the ACS/WFC F606W filter (left) and the ACS/SBC F150LP filter (right). Each panel spans $\sim13''\times13''$ (about $\sim 1/4$ of a single image), and the scale bar indicates $5''$. Two sources circled in blue are detected at high significance in the far-UV. The two circled sources have far-UV S/N of $\sim$19 (brighter) and $\sim$4 (fainter). Although the far-UV image contains fewer sources overall, it is very deep ($\sim28.7$ AB), depending on exposure time and proximity to dark glow; \(3\sigma\) within a \(0.5''\) diameter aperture). Centroid positions were determined for each high-S/N, compact source. A comparison of the pixel positions of source centers in the two images is used to measure the pre-alignment offsets, which are typically $\sim1''$ in translation and $<1^\circ$ in rotation.
}\label{fig:offset}\end{figure}

After stacking of individual pointings, we identify compact, high signal-to-noise (S/N) sources common to both the FUV and optical images. These sources serve as reference points for calculating and applying a rigid transformation (translation and rotation) (with unity scale) to align the SBC frames with the World Coordinate System (WCS) of the optical images, which are mutually registered and tied to a common Gaia-based astrometric frame. Geometric-distortion corrections are applied during the drizzling stage via the ACS/SBC distortion solution, so the remaining relative registration is well described by a rigid transform. See Figure~\ref{fig:offset} for examples.
 
Given a set of bright sources identified in the SBC image, we measure their initial projected positions \((x'_i, y'_i)\) in the optical reference frame by transforming the SBC pixel coordinates to sky coordinates using the SBC WCS, and then transforming those sky coordinates into optical-mosaic pixel coordinates using the optical mosaic WCS, as recorded in the image headers. However, because of uncertainties in the absolute pointing and roll angle of {\it Hubble}, the projected source positions do not coincide exactly with their optical counterparts.

For each source, we search in the vicinity of \((x'_i, y'_i)\) to identify the true counterpart \((x_i, y_i)\), typically associated with a nearby bright source in the optical image. We then model the transformation between the two frames as a rigid 2D rotation about a fixed pivot point \((x_p, y_p)\), followed by a translation. This pivot point is not arbitrarily chosen; it is given in the header of the optical image and typically lies near the center of the field.
We define relative coordinates with respect to the pivot point as $x_{\text{rel}}=x_i-x_p$ and $y_{\text{rel}}=y_i-y_p$.

\begin{equation}
\begin{pmatrix}
x_{\text{trans}}\\
y_{\text{trans}}
\end{pmatrix}
=
\begin{pmatrix}
x_p\\
y_p
\end{pmatrix}
+
\begin{pmatrix}
\Delta x\\
\Delta y
\end{pmatrix}
+
\begin{pmatrix}
\cos\theta & -\sin\theta\\
\sin\theta & \cos\theta
\end{pmatrix}
\begin{pmatrix}
x_{\text{rel}}\\
y_{\text{rel}}
\end{pmatrix}.
\end{equation}

Here, $\Delta x$ and $\Delta y$ are translation offsets, and $\theta$ is the rotation angle. 

To find the optimal alignment, we minimize the sum of squared residuals between transformed source coordinates and the corresponding positions in the optical reference frame \((x'_i, y'_i)\):
\begin{equation}
E(\Delta x, \Delta y, \theta) = \sum_i \left[ (x_{\text{trans}, i} - x_i')^2 + (y_{\text{trans}, i} - y_i')^2 \right]
\label{eq:align_obj}
\end{equation}

We solve \eqref{eq:align_obj} using SciPy’s trust-region optimizer \texttt{scipy.optimize.minimize}, \texttt{method=`trust-constr'} on $(\Delta x,\Delta y,\theta)$. The solver takes cautious, size-limited steps and accepts an update only if it reduces $E$; otherwise, it shortens the step. This prevents unstable jumps and yields reliable convergence even with a few matches.

Typical alignment corrections are on the order of $\sim1''$ in translation and less than $1^\circ$ in rotation.
The remaining residuals after alignment are typically $\sim0.03\arcsec$, ensuring that any observed UV-to-optical offset greater than this threshold likely reflects physical differences rather than alignment errors. This level of precision allows users to confidently compare UV and optical features of galaxies down to the pixel level.

Once the best-fit transformation parameters are computed, we update the WCS headers of the individual SBC exposures and re-drizzle the data to align them with the optical mosaics. This ensures that the astrometric calibration is consistent across wavelengths.

Alignment was performed on all SBC-F150LP images from the GOODS-S, GOODS-N, and COSMOS fields. 12 of our 151 pointings lie outside the coverage of any optical image. While two sources are sufficient in principle to solve for a shift and rotation, in practice, we require more than two reliable matches for alignment. Due to the limited number of suitable sources and the small SBC field of view, alignment was not always successful. Of the 139 SBC pointings overlapping with optical imaging, 127 were successfully aligned. The success rate by field is:

\begin{itemize}
  \item \textbf{GOODS-South:} 70 out of 73 pointings
  \item \textbf{GOODS-North:} 29 out of 30 pointings
  \item \textbf{COSMOS:} 28 out of 36 pointings
\end{itemize}

The remaining 12 pointings lacked a sufficient number of reliable sources for an accurate transformation.

The lower success rate in COSMOS is largely due to the shallower depth of its SBC exposures (many received only a single orbit, Figure~\ref{fig:depthhist}), which limited the number of high-S/N detections.

Finally, the SBC/F150LP images are drizzled onto the optical mosaic grid (to enable consistent photometry and catalog construction) at $0.06''$/pixel for the GOODS fields and $0.03''$/pixel for COSMOS.

\subsection{Dark “glow” model \label{subsec:Dark Glow Model}}

The SBC camera uses a Multi-Anode Microchannel Array (MAMA) detector. Unlike CCDs, MAMA detectors do not exhibit read noise and are less affected by cosmic rays, making them well-suited for low-background far-UV observations. According to \citet{2017acs..rept....4A}, the detector temperature gradually increases during observations, and once it approaches approximately 25°C, the central region of the detector exhibits an increase in dark current. Below this threshold, the dark current remains low and spatially uniform, with an average value of $8.11 \times 10^{-6}\,\mathrm{cts\,s^{-1}\,pix^{-1}}$. This analysis was conducted using dark frames obtained when the telescope shutter was closed, ensuring that the measured dark current originates purely from the instrument itself and not from any celestial or sky background. Based on these characteristics, the dark current in the SBC detector can be modeled using two components: (1) a uniform baseline component that dominates when the temperature is below 25°C, and (2) an additional, spatially non-uniform ``glow'' centered near the detector that becomes evident when the instrument warms above 25°C.

To characterize the spatial shape of this central dark ``glow'', we construct a stacked background from all SBC \texttt{FLT} exposures (the calibrated, flat-fielded \texttt{flt.fits} files produced by the standard \textit{HST} pipeline). Because any astrophysical flux would bias that stack, we first need to identify and exclude source pixels across the field before combining frames. We begin by creating a segmentation map using the \texttt{Photutils} package, applied to a high-resolution optical image such as F606W from the Hubble Legacy Fields (HLF; \citealt{2016arXiv160600841I, Whitaker_2019}). This map identifies and labels all detected stars and galaxies in the field. We choose an optical image as the reference for segmentation, rather than a UV image, because the optical data are generally much deeper, detecting more galaxies and the fainter, outer regions of the galaxies. Many sources that are clearly detected in the optical may be faint or completely undetected in the UV due to intrinsic spectral properties, dust attenuation, or redshift effects. Using the optical image as the basis, therefore, ensures a more complete segmentation map.

Next, we reverse-drizzle (i.e., “blot”) this segmentation map onto each FLT frame using \texttt{DrizzlePac}, aligning it with the native detector geometry. This allows us to accurately mask the positions of known sources in every FLT exposure. After masking, we subtract the uniform dark current by multiplying the uniform dark rate by the exposure time. 

To construct a precise model of the detector’s dark glow, we co-add more than 1000 FLT images after the uniform dark current has been subtracted and all visible sources have been masked. For each location in the detector, we accumulate the total counts across all images, using only the unmasked regions. Pixels that are masked in a given FLT (due to the presence of galaxies or other objects) are excluded from both the count summation and the exposure time accumulation. This yields a stacked image of counts together with an exposure map that represents the total unmasked exposure time per pixel. Dividing the total counts by the cumulative exposure time produces a mean image in units of counts per second. To further refine this result, we identify and exclude bad and hot pixels that arise from detector artifacts and would otherwise contaminate the smooth background, using the standard ACS/SBC bad-pixel table (BPXTAB). Finally, we fit a two-dimensional ninth-order polynomial to the cleaned image,
using a least-squares minimization with the Levenberg--Marquardt routine
\texttt{scipy.optimize.leastsq} to model the spatial structure of the dark glow
(selected after testing multiple orders to minimize residual structure). This high-order fit produces a smooth master dark glow model, which significantly improves background subtraction and enhances the reliability of subsequent photometric measurements. For context, Figure~5 of \citealt{2017acs..rept....4A} shows that the dark rate in the central elevated region is $\sim2\times$ higher than in a low-rate corner region at $T\sim25^\circ$C, and increases rapidly at higher temperatures.

\begin{figure*}
[htb!]\centering\includegraphics[width=1\textwidth]{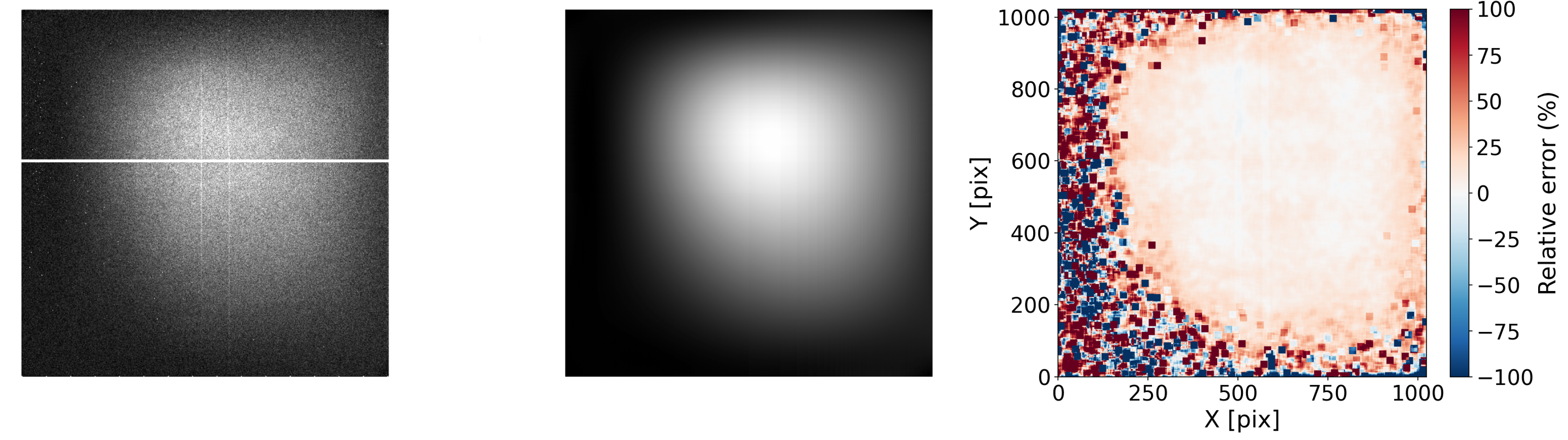}\caption{
\textbf{Left:} Stacked ACS/SBC exposures (uniform dark current subtracted and sources masked), revealing the residual background from detector glow plus diffuse sky; bad and hot pixels are excluded.
\textbf{Middle:} Ninth-order 2D polynomial fit to the left panel, providing a smooth master dark glow model. In both the stacked (left) and the model (middle), the maximum of the dark glow is \(\approx15\times\) the level of diffuse sky background around the edges of the frame. 
\textbf{Right:} Pixel-wise percent-difference map between the model (middle) and the data (left). For display, a NaN-aware \(20\times20\) boxcar smoothing is applied. This demonstrates that the model is tracking the shape of the dark glow to less than 7\% across peak of the dark glow signal.  
}\label{fig:model}\end{figure*}

Although our analysis does not explicitly focus on detector temperature, we examined a small subset of exposures, selecting only those taken at low detector temperatures, where only the uniform dark current component is expected to be present \citep{2017acs..rept....4A}. After masking all galaxies and subtracting the expected uniform dark current from these low-temperature exposures, we still observed an additional residual signal that is nearly uniform across the detector, with a mean value of approximately $6.5 \times 10^{-6}$ cts s$^{-1}$ pix$^{-1}$. The maximum dark glow signal is $\sim15\times$ higher than this level (Fig.~\ref{fig:model}). The presence of this residual signal, despite the removal of the uniform dark current and astrophysical sources, suggests a diffuse sky background origin. Possible contributors include airglow, zodiacal light, or Galactic emission (e.g., \citealt{2000AJ....120.1153B}).

While we do not attempt to characterize this background in detail, we remove its contribution on an image-by-image basis. To do this, we define a fixed region near the corner of each image—far from the central dark glow (the lower left corner in Figure \ref{fig:model}). For each FLT exposure, after subtracting the uniform dark current, we compute the mean signal in this corner region using pixels that have not been identified with bad pixels or optical sources. This mean value, representing the diffuse sky background in counts/s/pix, is multiplied by the exposure time of the FLT and subtracted from all pixels in that image. This per-image correction effectively removes the diffuse sky background. 

Furthermore, this constant level, whose origin will be investigated further in our future analysis, is also subtracted from our final dark glow model. As a result, the dark glow model approaches zero near the detector edges, where the detector glow contribution is minimal.

\subsection{Dark “glow” subtraction and stacking\label{subsec:Data Reduction and Image Processing}}
\label{sec:uvopt_examples_text}

To isolate the pure astrophysical signal from galaxies in the far-UV images, we subtract three distinct background components from each FLT exposure: the uniform dark current, the diffuse sky background, and the dark “glow” pattern.

In order to subtract the dark glow, we need to know its amplitude in each exposure. To determine the amplitude, we measure the average signal within a circular region (radius of 200 detector pixels) centered on the peak of the glow signal, after subtracting the uniform dark current and the diffuse sky background, and excluding pixels identified with sources as in previous steps (see Section \ref{subsec:Dark Glow Model}). In parallel, we compute the total signal from the dark glow model in the same region (with the same masks applied). The ratio of these two sums yields the per-image scale factor for the dark glow model.

We then subtract the three background components as follows:
\begin{itemize}
  \item The uniform dark current, scaled by the image exposure time
  \item The diffuse sky background, scaled to the remaining signal in image corner, with sources masked
  \item The dark glow model, scaled to the remaining signal near the glow peak, with sources masked.
\end{itemize}

The result is a fully background-subtracted image that contains only signal from stars and galaxies. Because many pixels detect zero photons, many of the dark-subtracted pixels in the final image are negative.

In addition to producing the background-subtracted FLTs, we need to determine the total background signal contributed to each output pixel in the final mosaic. To do this, we drizzle the three background components—uniform dark, diffuse sky, and scaled dark glow—onto the common output frame. This allows us to reconstruct the total background per pixel and generate a variance map and an exposure time map for the mosaic. Together with the segmentation map, these products—the fully dark-subtracted mosaic, the variance map, and the exposure time map—enable photometry and catalog construction.

\begin{figure}
    \centering
    \includegraphics[width=\linewidth]{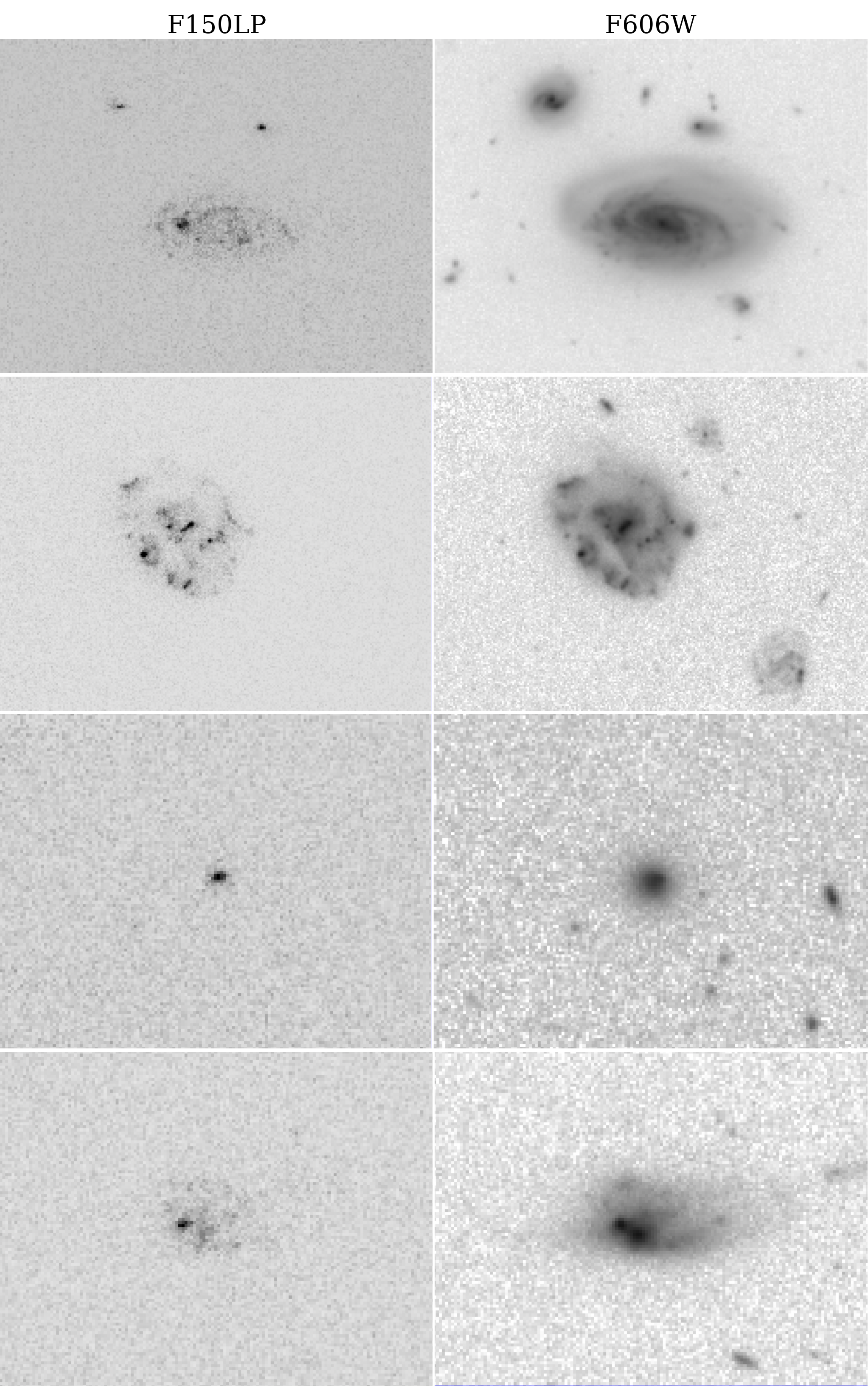}
    \caption{
    Matched cutouts in F150LP (left) and F606W (right).
    From top to bottom: two spirals, an elliptical, and an irregular galaxy.}
    \label{fig:uvopt_examples}
\end{figure}

Figure~\ref{fig:uvopt_examples} shows example images of bright galaxies (not representative of our typical detection) in both F150LP (far-UV) and F606W (optical). 
In the far-UV, spiral morphologies resolve into patchy star-forming knots; the elliptical galaxy is 
faint but clearly detected, and the irregular galaxy presents compact UV clumps. 
Many galaxies that are prominent in the optical are absent in F150LP because their far-UV emission is weak due to older stellar populations or dust-attenuation, or beyond $z\sim1$, where the Lyman limit shifts to the red side of the filter, and both galaxy and intergalactic neutral hydrogen can attenuate the ionizing photons \citep{2014MNRAS.442.1805I}.

\section{Source Detection and Photometry}
\label{sec: Source Detection and Photometry}

\subsection{Segmentation Map Construction}
To accurately detect sources and generate the photometric catalog, we begin by constructing a segmentation map that defines the pixels associated with each star or galaxy. At first, we considered using previous optical segmentation maps like that of \citet{Whitaker_2019}, but those maps employed depth-dependent detection thresholds. Specifically, deeper regions such as the HUDF \citep{2013ApJS..209....6I} had large segmentation areas down to very low surface brightness, while the shallower regions from GEMS \citep{2004ApJS..152..163R} data would have smaller areas. To minimize field-to-field selection biases, we impose a uniform surface-brightness detection criterion across all fields. For the F606W reference mosaics, we adopt $26.14~\mathrm{AB~mag~arcsec^{-2}}$, and for the F814W reference mosaics, we adopt $25.82~\mathrm{AB~mag~arcsec^{-2}}$; in each case, the threshold is derived from the corresponding mosaic zeropoint and the adopted pixel scale and then converted to the equivalent per-pixel threshold used for source detection. The small offset between the F606W and F814W thresholds accounts for modest differences in galaxy colors and filter throughput, and yields closely comparable detection behavior between the two reference datasets. This yields a segmentation map with a similar selection function across the field, which is crucial for consistent photometry.

Subsequently, we smooth the optical image using a 2-D Gaussian filter with a sigma of $0.12''$ (a modest smoothing scale chosen to suppress pixel-scale noise while preserving source morphology; implemented as $\sigma=2$ pixels for GOODS ($0.06''$/pix) and $\sigma=4$ pixels for COSMOS ($0.03''$/pix)) to reduce noise and enhance source detectability.
Following smoothing, we create the segmentation map using \texttt{Photutils} with a detection threshold area of at least $0.068~\mathrm{arcsec}^2$ corresponding to 19 connected pixels (where a connected pixel can be any of the 8 surrounding pixels) at $0.06''$/pixel above the uniform surface brightness threshold explained above. We then deblend closely spaced sources using 32 intensity levels and a contrast threshold of 0.001.

\subsection{Measurements and Aperture Definition}

We measure far-UV fluxes using the background-subtracted SBC/F150LP mosaics by integrating each source over its corresponding segmentation region. Denoting the segmentation map by $\mathcal{S}$ and the background-subtracted image by $I$, the total source count rate (in electrons s$^{-1}$) is
\begin{equation}
F \;=\; \sum_{i \in \mathcal{S}} I_i ,
\label{eq:seg_flux}
\end{equation}
where $I_i$ is the value of pixel $i$.

Because low-level residual structure can remain after the dark-glow subtraction, we apply a local background correction. For each source, we measure the per-pixel residual background in an annulus from $1.8''$ to $3''$ around the flux-weighted centroid, masking pixels that belong to other segmented objects, and subtract the corresponding total background from the aperture sum (i.e., this per-pixel level multiplied by the number of valid pixels in the aperture, equivalently subtracting it from each aperture pixel before summing) to obtain the net galaxy flux.
After subtracting the uniform dark current, diffuse sky, and the master dark-glow model, the residual background is expected to be consistent with zero away from sources. The annular measurement, therefore captures any local offset: a positive value indicates slight under-subtraction and is removed from the source flux, while a negative value indicates slight over-subtraction; subtracting a negative value correspondingly adds back the missing counts, bringing the local background toward zero.

\subsection{Photometric Error Estimation and Signal-to-Noise Criteria}

To estimate photometric uncertainties, we must first recover the effective exposure time corresponding to each pixel, even though the drizzled science images and variance maps are in electrons s$^{-1}$. The exposure-time maps produced by \texttt{drizzle} are scaled by the ratio of the output pixel area to the native SBC detector pixel area ($0.034''\times 0.030''$), and therefore do not represent the true integration time. Drizzling converts pixel values from electrons to electrons s$^{-1}$ and adjusts the exposure-time maps accordingly; however, changes in pixel scale and projection mean that the output exposure-time maps no longer equal the actual exposure time. Let the final drizzle produce an exposure-time map $T_{\mathrm{final}}$. If the final mosaic has pixel area $A_{\mathrm{final}}$ and the native SBC detector has pixel area $A_{\mathrm{native}}$, we scale the effective exposure time by the pixel-area ratio,
\begin{equation}
t_{\mathrm{scaled}} \;=\; T_{\mathrm{final}}\left(\frac{A_{\mathrm{native}}}{A_{\mathrm{final}}}\right)\, .
\end{equation}

Flux uncertainties are computed from the variance and (scaled) time maps within each source’s segmentation mask, assuming background-dominated Poisson statistics. Because the ACS/SBC MAMA is photon-counting and the drizzled count-rate image is quantized (often zero in low-background regions), we do not estimate uncertainties from the pixel signal itself; instead, we use the drizzle variance map derived from the modeled SBC background (uniform dark, diffuse sky, and dark glow).
Specifically, we multiply the variance map by the \emph{scaled} exposure time map, sum over pixels inside the segmentation mask, take the square root, and then divide by the mean scaled exposure time:
\begin{equation}
\Delta F \;=\; 
\frac{\sqrt{\sum\limits_{i \in \mathcal{S}} \left( \mathrm{Var}_i \times t_{\mathrm{scaled},i} \right)}}
{\overline{t}_{\mathrm{scaled}}} \, ,
\label{eq:deltaN}
\end{equation}
where $\mathcal{S}$ denotes the source segmentation region and $\overline{t}_{\mathrm{scaled}}$ is the mean of $t_{\mathrm{scaled},i}$ over that same region.

Intuitively, the quantity under the square root is the variance of the \emph{total electrons} collected across the segmentation region. The variance map supplies a per-second variance for each pixel; multiplying by the local (scaled) exposure time $t_{\mathrm{scaled},i}$ converts that to a variance in electrons, and summing over pixels adds those independent variances. Taking the square root gives the $1\sigma$ uncertainty in electrons. Dividing by the mean scaled exposure time $\overline{t}_{\mathrm{scaled}}$ then converts this back to electrons s$^{-1}$, i.e., the same units as the measured flux, so the quoted uncertainty is normalized consistently with the flux measurement.

In addition to isophotal photometry defined by the segmentation maps, we also perform circular-aperture photometry with radii of $0.3''$, $0.6''$, $1.2''$, and $1.8''$, in order to perform consistency checks and assess detection significance across multiple spatial scales. A source is considered detected if its net flux exceeds the $3\sigma$ threshold.

\subsection{Zeropoint and Count-Rate Conversion}

AB magnitudes are calculated using the ACS/SBC F150LP zeropoint of 22.76 for our observation epoch. This zeropoint decreased by 0.06 (6\%\ in flux) over the duration of these observations (from 2002 to 2016). We choose to use a single zeropoint chosen at the exposure-weighted year of 2007. Thus, there is an additional intrinsic error of $\sim3\%$ for any given flux in the catalog, which is about twice the reported absolute flux calibration error of 1.4\%\ \citep{2019acs..rept....5A}. 

Thus, a single, representative zeropoint is sufficient for the catalog.
 AB magnitudes are computed as
\begin{equation}
\mathrm{mag}_{\mathrm{AB}} \;=\; 22.76 \;-\; 2.5 \log_{10}\!\left(F_{\mathrm{corrected}}\right),
\label{eq:abmag}
\end{equation}
where $F_{\mathrm{corrected}}$ is the background- and aperture-corrected count rate in electrons per second.

Count rates are converted to physical flux densities using the instrument calibration factor $\mathrm{PHOTFLAM}=3.3225 \times 10^{-17}\,\mathrm{erg\,cm^{-2}\,s^{-1}\,\AA^{-1}}$. We apply the same conversion to the flux uncertainties derived in Equation~\ref{eq:deltaN}.

\subsection{Milky Way (Galactic) extinction}\label{sec:mw_ext}

All catalog photometry is corrected for foreground Milky Way extinction. We adopt the \citet{2011ApJ...737..103S} recalibration of the \citet{Fitzpatrick_1999} MW law with $R_V=3.1$. For ACS/SBC F150LP we use the pivot $\lambda_{\rm piv}=1605$\,\AA, giving $k_{1605}\!\equiv\!A_{1605}/E(B{-}V)=7.926$. Per-field reddenings $E(B{-}V)$ are taken from the IRSA DUST service, which reports the \citet{2011ApJ...737..103S} values at the field centers.\footnote{\url{https://irsa.ipac.caltech.edu/applications/DUST/}} We apply $A_{1605}=k_{1605}\,E(B{-}V)$ and thus $m_{\rm corr}=m_{\rm obs}-A_{1605}$ and $F_{\rm corr}=F_{\rm obs}\,10^{0.4A_{1605}}$. \emph{Uncertainties:} since the correction is multiplicative, we scale flux errors by the same factor $C\!=\!10^{0.4A_{1605}}$, so S/N is unchanged. Table~\ref{tab:mw_ext_1605} lists the values used in each field; all reported magnitudes and fluxes are already corrected.

\begin{deluxetable}{lcccc}
\tablecaption{Milky Way extinction at F150LP ($\lambda_{\rm piv}=1605$\,\AA) using \citet{2011ApJ...737..103S} $E(B{-}V)$ values.\label{tab:mw_ext_1605}}
\tablehead{
\colhead{Field} & \colhead{$E(B{-}V)$} & \colhead{$A_{1605}$ (mag)} & \colhead{Flux factor}
}
\startdata
GOODS-S & 0.0071 & 0.056 & 1.053 \\
GOODS-N & 0.0096 & 0.076 & 1.073 \\
COSMOS  & 0.0155 & 0.123 & 1.120 \\
\enddata
\tablecomments{Flux factor is $10^{0.4A_{1605}}$.}
\end{deluxetable}

\section{Catalog Construction and Description}
\label{sec:catalog-construction-and-description}

\subsection{Catalog Columns}

In GOODS-S, the HLF GOODS-S astrometric solution applies constant Gaia-frame offsets \citep{Whitaker_2019}. We find that the reported offset signs are reversed, and therefore we apply the corrected-sign offsets when constructing the Gaia-corrected coordinates. The resulting corrections are encoded in the RA Gaia and Dec Gaia columns via the equations given in Table~\ref{tab:catalog_columns}.

Our final catalog includes the columns summarized in Table~\ref{tab:catalog_columns}.

\begin{deluxetable*}{ll}
\tablewidth{0pt}
\tabletypesize{\footnotesize}
\tablecaption{Catalog Columns and Descriptions}
\label{tab:catalog_columns}
\tablehead{\colhead{Column} & \colhead{Description}}
\startdata
\parbox[t]{0.24\textwidth}{Galaxy ID} & \parbox[t]{0.72\textwidth}{Unique identifier corresponding to the label assigned to each detected object.} \\
\parbox[t]{0.24\textwidth}{CANDELS ID} & \parbox[t]{0.72\textwidth}{ID of the nearest CANDELS counterpart within $0.4''$ \citep{2013ApJS..207...24G,2017ApJS..228....7N,2019ApJS..243...22B}; set to $-99$ if none.} \\
\parbox[t]{0.24\textwidth}{3D-HST ID} & \parbox[t]{0.72\textwidth}{ID of the nearest 3D-HST counterpart within $0.4''$ \citep{2014ApJS..214...24S,2012ApJS..200...13B}; set to $-99$ if none.} \\
\parbox[t]{0.24\textwidth}{$x, y$} & \parbox[t]{0.72\textwidth}{Pixel coordinates of the flux-weighted centroid measured on the optical reference image (F606W and F814W).} \\
\parbox[t]{0.24\textwidth}{RA, Dec} & \parbox[t]{0.72\textwidth}{Right ascension and declination (J2000, degrees).} \\
\parbox[t]{0.24\textwidth}{RA Gaia} & \parbox[t]{0.72\textwidth}{R.A. (J2000, degrees) corrected to the Gaia DR2 astrometric frame by applying a constant WCS offset to the HLF GOODS-S coordinates \citep{Whitaker_2019}: $\mathrm{RA}_{\rm Gaia}=\mathrm{RA}-0.1130/3600~\mathrm{deg}$ (i.e., $-0.1130''$). Applied \emph{only} for the GOODS-South field.} \\
\parbox[t]{0.24\textwidth}{Dec Gaia} & \parbox[t]{0.72\textwidth}{Decl. (J2000, degrees) corrected to the Gaia DR2 astrometric frame by applying a constant WCS offset to the HLF GOODS-S coordinates \citep{Whitaker_2019}: $\mathrm{Dec}_{\rm Gaia}=\mathrm{Dec}+0.26/3600~\mathrm{deg}$ (i.e., $+0.26''$). Applied \emph{only} for the GOODS-South field.} \\
\parbox[t]{0.24\textwidth}{Npix} & \parbox[t]{0.72\textwidth}{Number of pixels within the segmentation region.} \\
\parbox[t]{0.24\textwidth}{Flux} & \parbox[t]{0.72\textwidth}{Flux density converted using \textsc{photflam} (erg cm$^{-2}$ s$^{-1}$ \AA$^{-1}$).} \\
\parbox[t]{0.24\textwidth}{Flux Error} & \parbox[t]{0.72\textwidth}{Uncertainty in physical units (erg cm$^{-2}$ s$^{-1}$ \AA$^{-1}$).} \\
\parbox[t]{0.24\textwidth}{Magnitude} & \parbox[t]{0.72\textwidth}{Total AB magnitude.} \\
\parbox[t]{0.24\textwidth}{Circular Aperture Measurements} & \parbox[t]{0.72\textwidth}{For each radius (e.g., $0.3''$, $0.6''$, $1.2''$, $1.8''$), we report: Flux, Flux Error, and AB Magnitude.} \\
\parbox[t]{0.24\textwidth}{Coverage flag} & \parbox[t]{0.72\textwidth}{\texttt{0} if the source's segmentation region touches the image boundary (i.e., part of the object lies outside the field), otherwise \texttt{1}.} \\
\enddata
\tablecomments{The Gaia-frame offsets for GOODS-S correct the HLF GOODS-S WCS to the Gaia DR2 frame. Use of the Gaia-corrected columns (RA Gaia, Dec Gaia) is recommended for cross-survey astrometry within GOODS-S.}
\end{deluxetable*}

\subsection{Coverage flag}\label{sec:coverage}

We provide a single per–object coverage flag (\texttt{COV}) indicating whether the full segmentation footprint of the source lies within the FUV image. We identify out-of-frame / NaN pixels in the variance map, dilate that mask by a small buffer (\texttt{PIXEL\_BUFFER} = 3 pixels, $\approx 0.18''$ at $0.06''$~pix$^{-1}$) to guard against sources whose segmentation footprint would be partially missing near regions of invalid FUV coverage (NaN variance pixels), and test whether the source footprint intersects this dilated NaN (edge-proximity) mask.

In the released catalog, \texttt{COV} is binary: \texttt{COV} = 1 for fully covered sources and \texttt{COV} = 0 for sources that touch the edge-proximity mask.

\subsection{Flux Losses and Aperture Recommendation}

The point-spread function of ACS/SBC FUV images is similar to ACS/WFC images, though very slightly broader. For example, the encircled energy in a $0.3''$ radius circular aperture is 0.885 with F606W in ACS/WFC and only 0.715 in F150LP (see ACS Instrument Handbook Sec.\ 5.6, Fig.\ 5.14)\footnote{\url{https://hst-docs.stsci.edu/acsihb/chapter-5-imaging/5-6-acs-point-spread-functions}}. We therefore did not do any convolutions to match the point spread functions between our optical detection image and the SBC FUV image. Fluxes of compact sources measured in small apertures can be underestimated considerably. To mitigate this, we recommend a minimum aperture area when measuring fluxes of very compact objects (\citealt{Melinder_2023}).

We recommend using the segmentation-based \texttt{Flux} for most sources. However, if the segmentation area is smaller than a circle of radius $0.3''$, then adopt the corresponding flux of the $0.3''$ radius circular aperture and convert to an approximate total flux using the ACS/SBC encircled-energy correction\footnote{\url{https://www.stsci.edu/hst/instrumentation/acs/data-analysis/aperture-corrections}.}. For ACS/SBC F150LP, the encircled-energy fractions are $\mathrm{EE}(0.3'')=0.715$, $\mathrm{EE}(0.6'')=0.848$, $\mathrm{EE}(1.2'')=0.933$, and $\mathrm{EE}(1.8'')=0.962$; in general, use $f_{\rm tot}=f_{\rm circ}/\mathrm{EE}(r)$. Thus, for $r=0.3''$ we adopt

\begin{equation}
f_{\rm tot} \;=\; \frac{1}{0.715}\; f_{\rm circ}\, .
\label{eq:correction}
\end{equation}

Conceptually, Equation~\ref{eq:correction} assumes a point source centered within the circular aperture. For resolved galaxies, the point-source encircled-energy correction is a lower limit to the true flux because significant light can lie beyond $r \approx 0.3''$. For users prioritizing maximal flux recovery, use fluxes in larger circular apertures (e.g., radii of $0.6''$ or $1.2''$) and verify convergence with curves of growth.

The complete catalog is publicly available through the Mikulski Archive for Space Telescopes (MAST) as a High-Level Science Product (HLSP; \dataset[FUEL-SURVEY]{\doi{10.17909/2yqw-3g14}}).

\section{Results}
\label{sec: Results}

\subsection{Sanity check of dark subtraction}
\label{sec:dark_sanity_grism}

To test that our dark modeling and subtraction do not introduce systematic offsets in the measured F150LP net flux (i.e., an additive residual background level), we use galaxies from the \textsc{3D-HST} catalogs \citet{Momcheva_2016} with redshifts in the range $1.7<z<4$ that fall within our F150LP coverage. At these redshifts, the F150LP filter (pivot wavelength $\lambda_p = 1606\,\mathrm{\AA}$; corresponding to rest-frame $\lambda_{\rm rest}\approx 320$--$600\,\mathrm{\AA}$ for $1.7<z<4$) samples the Lyman continuum of galaxies, which has been shown to be faint, with few detections at this redshift \citep{2007ApJ...668...62S, Siana_2010}. (Furthermore, the IGM should typically be opaque to photons of these energies at these redshifts \citep{2007ApJ...668...62S, 2014MNRAS.442.1805I}. We cross-matched the sources’ positions to our F150LP catalog using Astropy’s \texttt{match\_coordinates\_sky} routine, enforcing an angular-separation criterion of $<\!0.4\arcsec$. For each matched object, we
compute the aperture signal-to-noise ratio
$\mathrm{SNR_{aperture}} \equiv F_{\rm ap}/\sigma_{\rm ap}$
from the catalog aperture fluxes.

\begin{figure}
    \centering
    \includegraphics[width=\linewidth]{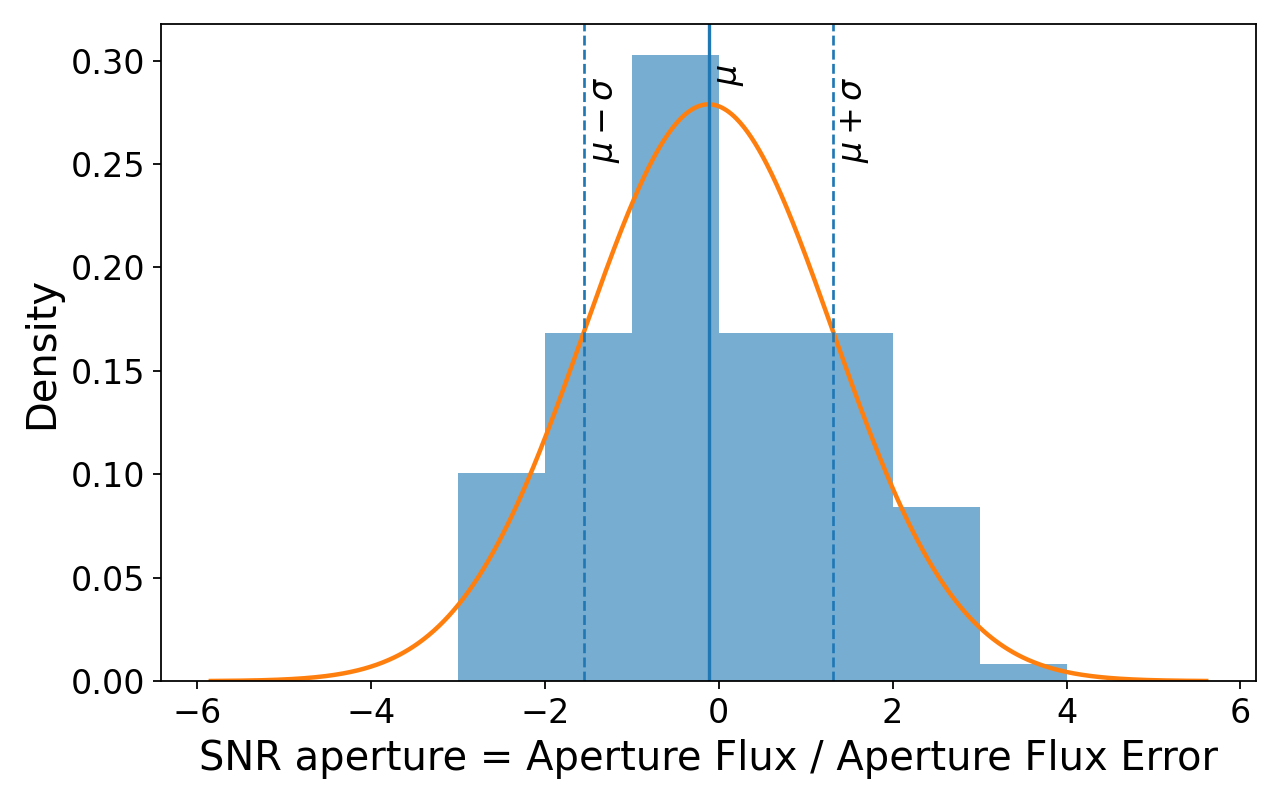}
    \caption{Distribution of $\mathrm{SNR_{aperture}} = F_{\rm ap}/\sigma_{\rm ap}$ for
    galaxies at $1.7<z<4$ matched within $<\!0.4\arcsec$ across GOODS--N/S and COSMOS.
    The curve shows the best-fit normal distribution; vertical lines mark $\mu$ and $\mu\pm\sigma$.
    We measure $\mu=-0.117\pm0.131$ and $\sigma=1.431\pm0.093$ for $N=119$, consistent with no additive offset (mean $\approx0$) but with a broader-than-unity dispersion.}
    \label{fig:snr}
\end{figure}

If backgrounds are unbiased and uncertainties are correctly calculated, the ensemble
$\mathrm{SNR_{aperture}}$ distribution should be consistent with a mean of zero and a standard deviation of one.
Combining GOODS--N/S and COSMOS, we obtain $N=119$ sources with a Gaussian fit ($\mu=-0.117$, $\sigma=1.431$; Fig.~\ref{fig:snr}).
The mean is consistent with zero within its uncertainty ($\mu=-0.117\pm0.131$), indicating no evidence for an additive residual background after dark subtraction.
The dispersion is broader than unity ($\sigma=1.431\pm0.093$), suggesting that the formal uncertainties are underestimated by a factor of $\approx1.43$, perhaps due to imperfect background subtraction. Due to this increased uncertainty, we caution users that galaxies in the catalog with statistical 3 sigma detections may not be real.  We thus think that a $4\sigma$ statistical significance (roughly $3\sigma \times 1.4$) is preferred for robust detections.

\subsection{External comparison with GALEX/FUV and HST/ACS-SBC HDF-N FUV imaging}
\label{subsec:ext_comp_f150lp}

We compared our SBC/F150LP photometry to GALEX/FUV \citet{2005ApJ...619L...1M} in COSMOS, GOODS--N, and GOODS--S, and to the \citet{Teplitz2006} F150LP catalog in GOODS--N.
For GALEX, we used the GR6 Deep Imaging Survey (DIS) data products accessed via the MAST GALEX GR6 interface.
Since \citet{Teplitz2006} predates the 2019 SBC calibration \citep{2019acs..rept....5A}, we added a constant $+0.22$\,mag to their magnitudes before the comparison.
Sources were matched within $0.5''$, we required $\mathrm{SNR}>4$ in our FUV sources for comparison.
The GALEX FUV PSF is broad (FWHM $\simeq 4.2''$), so this small match radius preferentially selects isolated, unambiguous counterparts.

We define 
$\Delta \equiv m_{\rm cmp}-m_{\rm SBC}$, where $m_{\rm cmp}$ is the magnitude taken from the external comparison dataset and $m_{\rm SBC}$ is our SBC/F150LP magnitude.
We report the median offset $\Delta_{\rm med}$.

\begin{figure}[t]
  \centering
  \includegraphics[width=\linewidth]{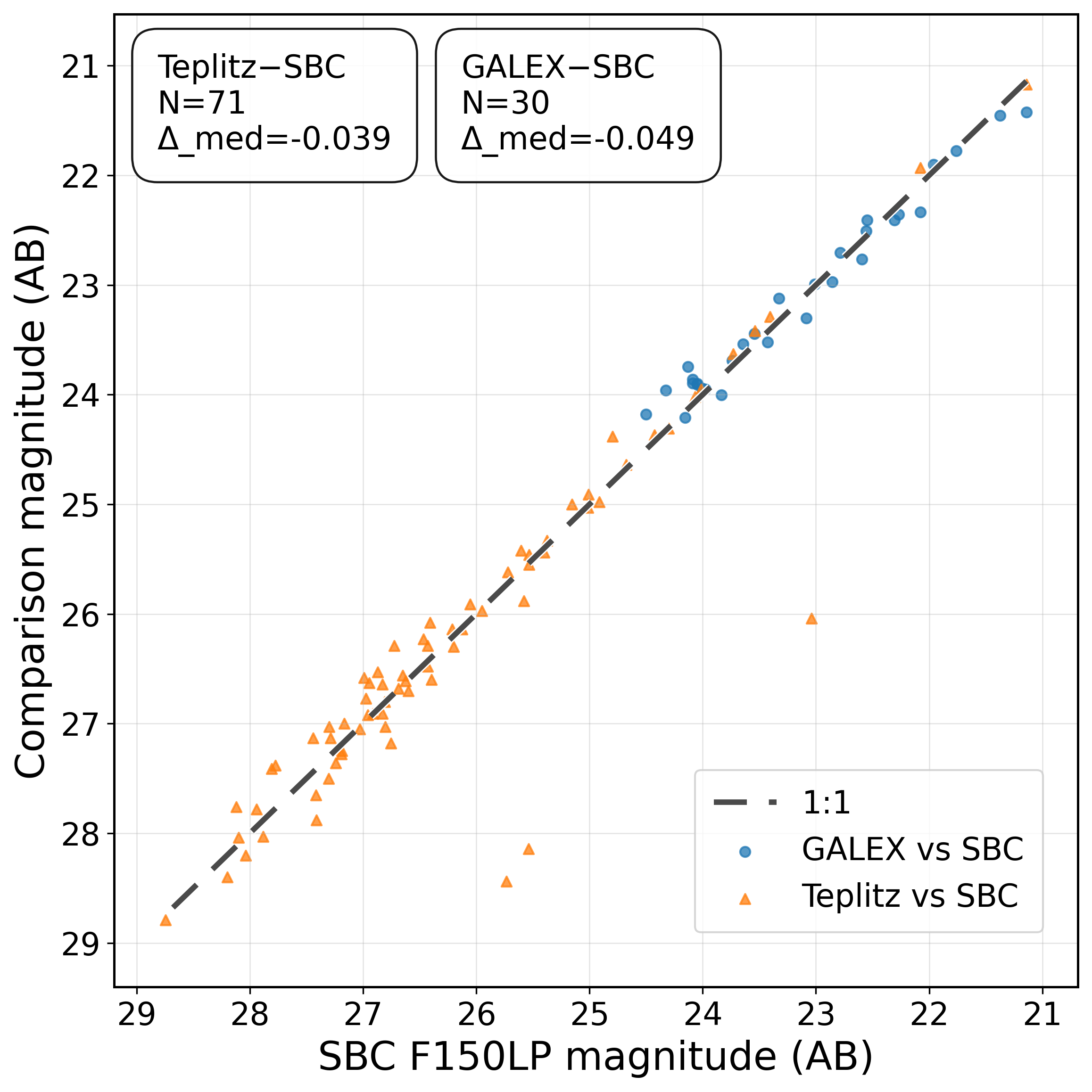}
  \caption{SBC/F150LP photometry compared to GALEX/FUV (blue circles) and to corrected \citet{Teplitz2006} F150LP (orange triangles) photometry, after applying a $+0.22$\,mag shift to the Teplitz magnitudes. The dashed line shows the $1{:}1$ relation.}
  \label{fig:galex_teplitz_sbc}
\end{figure}

In Fig.~\ref{fig:galex_teplitz_sbc}, we compare the photometry of these matched catalogs. After applying $+0.22$\,mag to \citet{Teplitz2006}, the Teplitz--SBC comparison gives the number of matched sources,
$N=71$, $\Delta_{\rm med}=-0.039$\,mag.
The GALEX--SBC comparison gives
$N=30$, $\Delta_{\rm med}=-0.049$\,mag.
These values are consistent with the $1{:}1$ relation with small median offsets.

Our segmentation areas are defined on the HST/ACS F814W image, whereas \citet{Teplitz2006} constructed a $V{+}I$ (F606W+F814W) segmentation map with a relatively high ($3.25\sigma$) threshold, resulting in smaller segmentation footprints. They then applied an aperture correction based on the $B$-band (F450W) flux difference between the $3.25\sigma$ and $0.65\sigma$ isophotal areas. In contrast, our aperture corrections assume a point source, so the remaining $\sim$5\% ($\sim$0.05\,mag) offset is likely driven by differences in (possibly incomplete) aperture corrections rather than by the choice of detection band alone.
We visually inspected the few most discrepant matches ($|\Delta|>0.5$\,mag) and confirmed they are consistent with aperture-definition differences.

\subsection{Survey Depth and Effective Area}

\begin{figure}
    \centering
    \includegraphics[width=\linewidth]{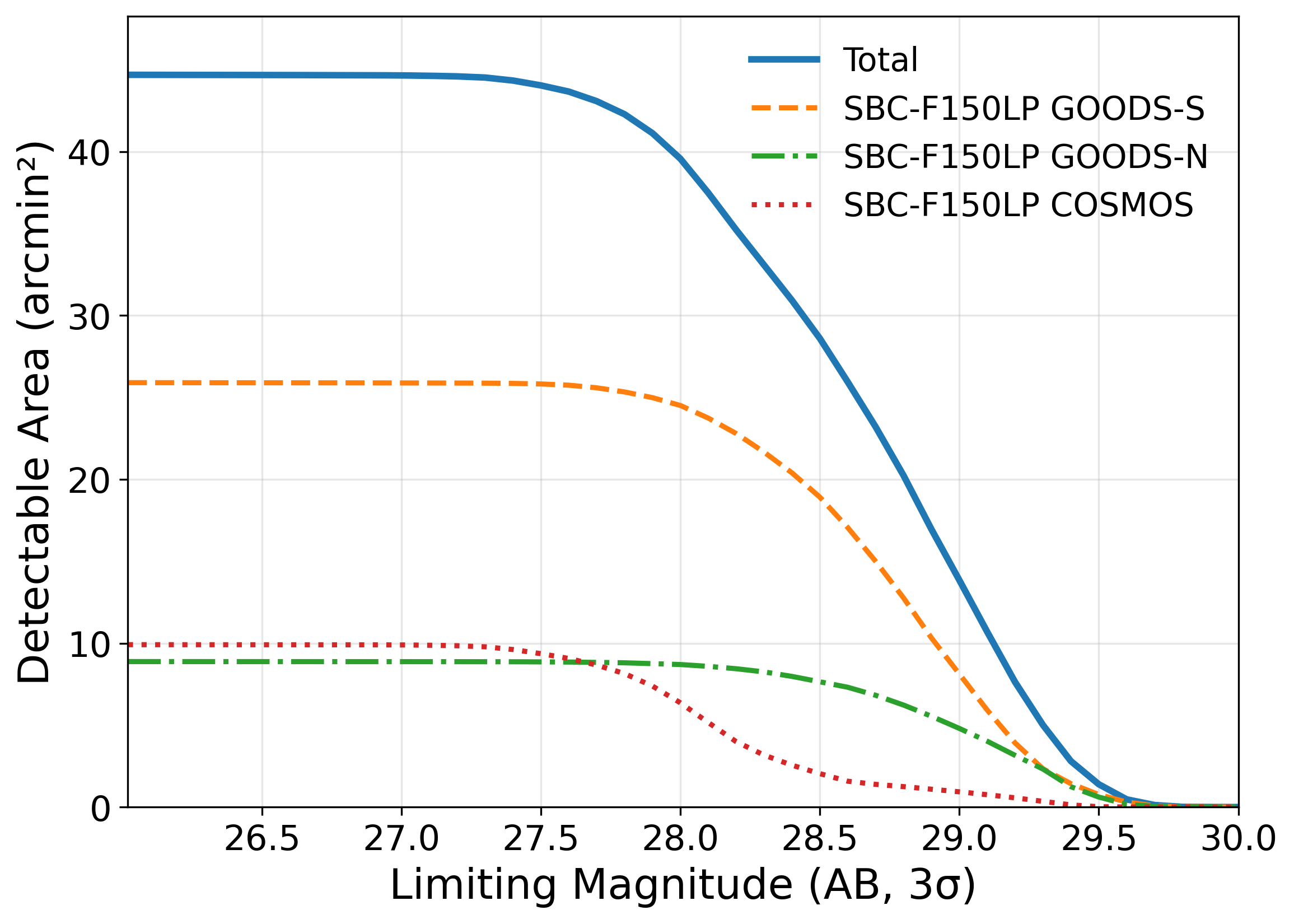}
    \caption{
        Detectable area as a function of limiting magnitude (AB, $3\sigma$) within a $0.5"$ diameter aperture for the SBC/F150LP mosaics. Curves are shown for GOODS-S (\textit{dashed}), GOODS-N (\textit{dot--dashed}), and COSMOS (\textit{dotted}), along with the \textit{solid} curve for the sum across fields. The curves are derived from per-pixel limiting-magnitude maps computed from exposure-time and variance/weight maps.
    }
    \label{fig:area_vs_mag}
\end{figure}

Imaging depth varies substantially across each SBC mosaic because the effective exposure time is not uniform and because the temperature-dependent dark-current ``glow'' varies across the detector. To capture these position-dependent sensitivities, we construct per-pixel weight/variance maps that propagate the effects of exposure time and the dark current following \citet{2007ApJ...668...62S}. From these variance maps, we derive a per-pixel $3\sigma$ limiting-magnitude map for a $0.5''$ diameter aperture. We then compute the cumulative detectable area as a function of limiting magnitude by integrating the sky area over which a source would meet S/N$=3$ in that aperture (see \citet{2007ApJ...668...62S} for a closely related implementation and their Fig.~6 for an early example of area–depth curves), as shown in Fig.~\ref{fig:area_vs_mag}.

The resulting curves are flat at bright magnitudes, where we are complete in all areas with coverage, and slowly decrease fainter than $\sim27.3$ (AB), where a smaller fraction of the mosaics have such high sensitivity. GOODS-S contributes the largest area at bright–intermediate depths, GOODS-N extends slightly deeper but over less area, and COSMOS adds a similar amount of area as GOODS-N, but mostly at shallower depths. Across all fields, we reach $m_{\mathrm{AB}}=27.3\ (3\sigma)$ in a $0.5''$ aperture over the entire footprint (44.7 arcmin$^2$), $\approx28.7$ over roughly half the area, and by $m_{\mathrm{AB}}\approx29.7$, the available area in these apertures is negligible.

\subsection{Redshift Distribution of SBC–3D-HST Matches}\label{sec:redshift-hist}

\begin{figure}
  \centering
  \includegraphics[width=\linewidth]{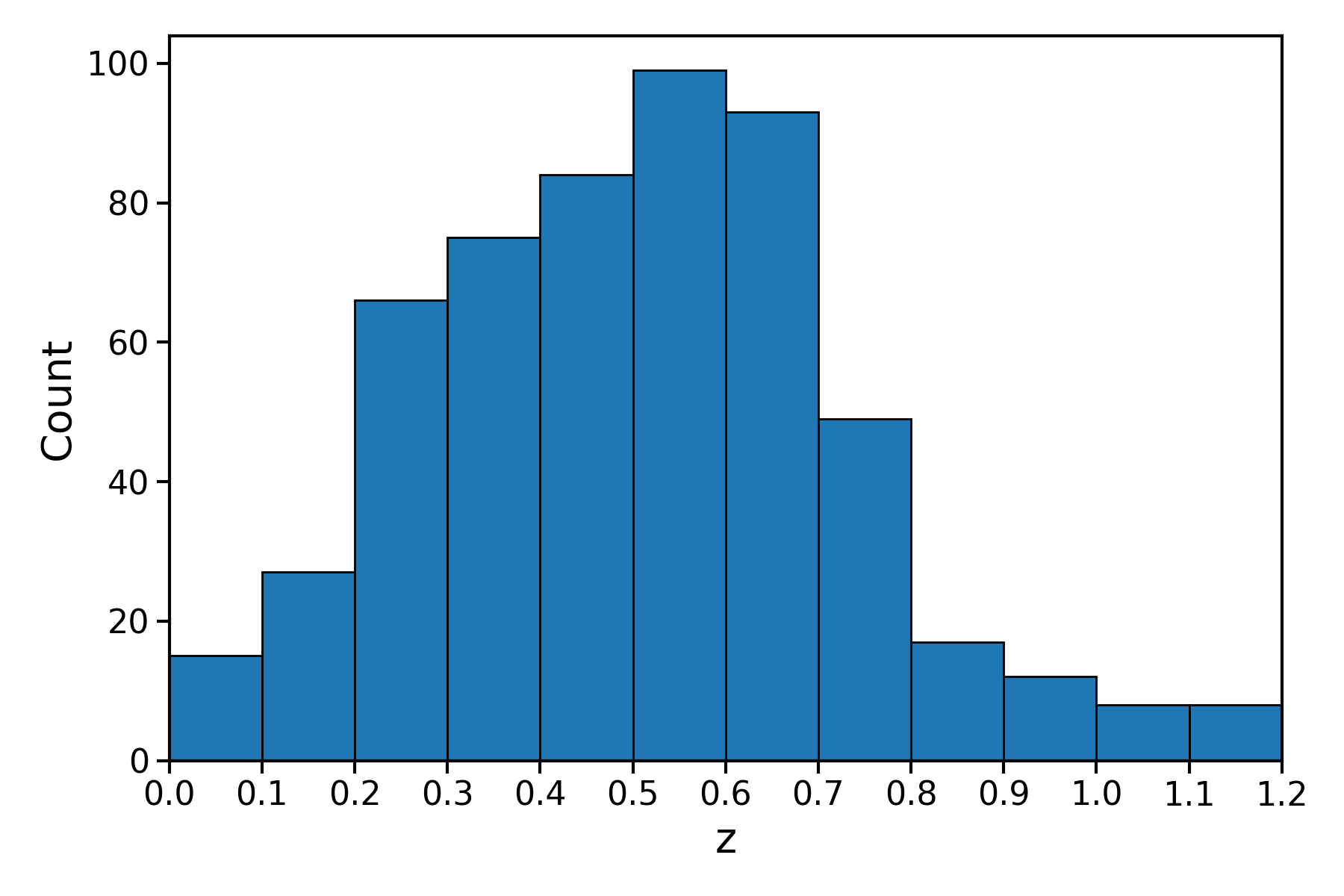}
  \caption{Redshift histogram of SBC–3D-HST matches in GOODS–N/S and COSMOS,
  $\Delta z=0.1$, $0\le z\le1.2$, and $\mathrm{SNR_{FUV}}\ge3$. The decline beyond $z\approx0.6$
  occurs as the Lyman break enters F150LP.}
  \label{fig:z_hist}
\end{figure}

Figure~\ref{fig:z_hist} shows the redshift distribution for 3D-HST matches to FUV detections in GOODS-N, GOODS-S, and COSMOS. We plot only sources with $\mathrm{SNR_{FUV}}\ge3$ and restrict to $0\le z\le1.2$ using $\Delta z=0.1$ bins. Counts rise to a peak near $z\approx0.5$–0.6 and then decline toward $z\approx1.2$.

The trend follows from the F150LP bandpass and the Lyman break. F150LP is a long-pass filter with a short-wavelength cutoff near $1500$\,\AA. As $z$ increases, the 912\,\AA\ Lyman break redshifts into the filter. At $z\approx0.7$, the break reaches the peak of the filter transmission, decreasing signal, and resulting in fewer detections. By $z\gtrsim1.2$, the band probes only the Lyman continuum, so typical galaxies contribute little flux.

We adopt the \textsc{3D-HST} best-estimate redshift $z_{\rm best}$ \citep{Momcheva_2016}, which uses spectroscopic redshifts when available, otherwise \textsc{3D-HST} grism redshifts, and photometric redshifts for the remaining sources.
The few claimed detections of LyC at $z>1.2$ require additional scrutiny and will therefore be left for a future analysis. We therefore limit the plot to $z\le1.2$.

\subsection{Science, Variance, and Exposure-Time Cutouts}
\label{subsec:variance_exptime_maps}

\begin{figure*}[t!]
\centering
\includegraphics[width=0.95\textwidth]{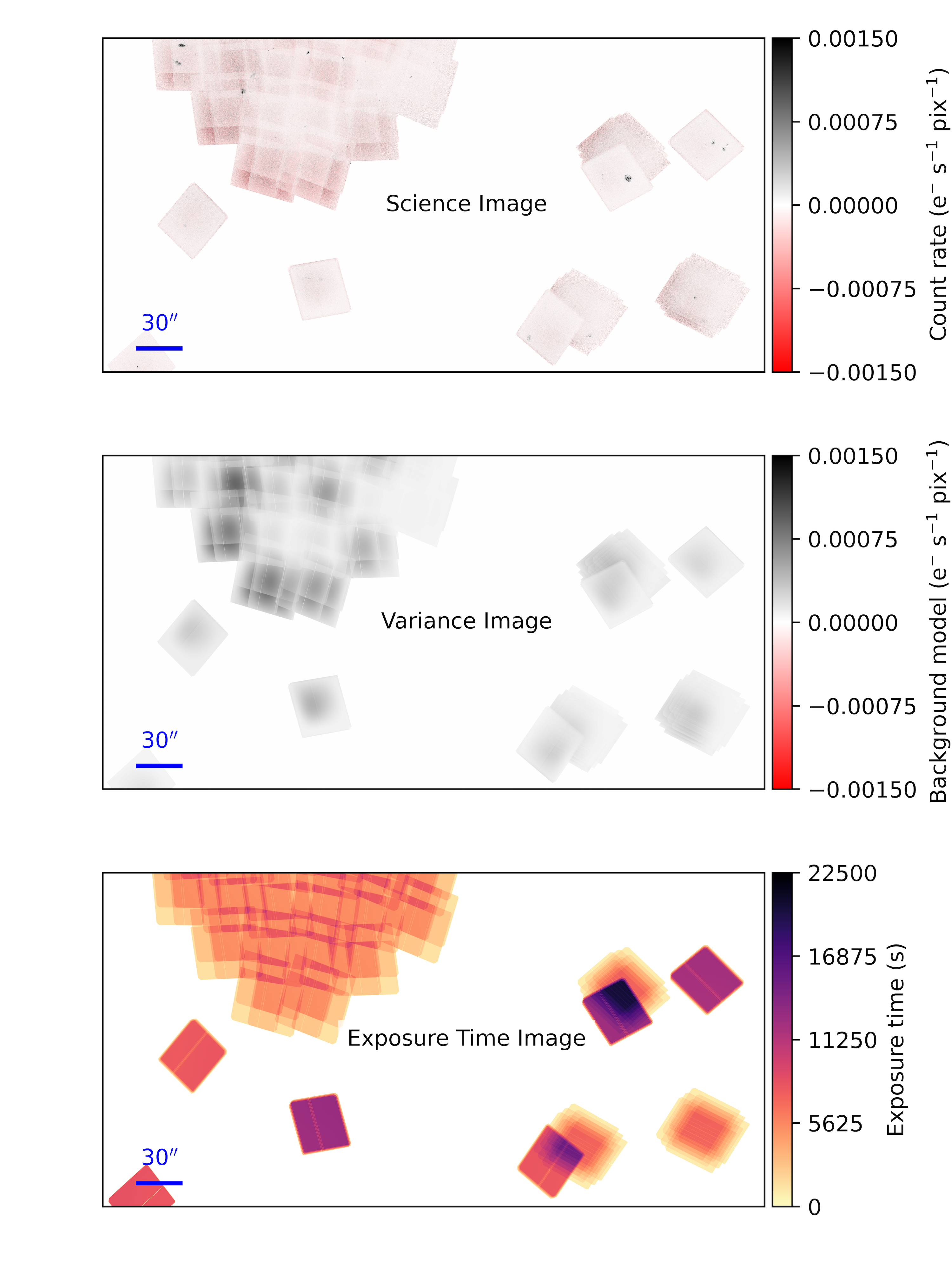}
\caption{Example F150LP maps cutout of a portion of the GOODS-S field. The area on the top left with many contiguous pointings is the UDF data from \citep{Teplitz2006}, while the other individual pointings are deep exposures from other programs. \textit{Top:} drizzled science mosaic in count-rate units (e$^{-}$\,s$^{-1}$\,pix$^{-1}$). \textit{Middle:} variance map shown on the same symmetric stretch to highlight spatial variations (the variance values are non-negative; negative tick labels reflect the display limits only). \textit{Bottom:} exposure-time map (colorbar in seconds. A 30\arcsec\ scale bar is shown in each panel.}
\label{fig:cutout}
\end{figure*}

Figure~\ref{fig:cutout} shows an example GOODS-S cutout illustrating the dispersed ACS/SBC F150LP coverage: the drizzled science mosaic (count rate), the corresponding variance map, and exposure-time map derived from the drizzled weight image (colorbar in seconds after applying the scale factor). While in the UDF, the exposure time is relatively uniform, the variance map still shows a dramatic spatial difference. We also note that, even though the mean background of the sky-subtracted science image is approximately zero, many pixels are negative, as expected from dark subtraction and reflected in the science image.

\section{Scientific Applications}
\label{sec:Applications}

The uniformly processed ACS/SBC F150LP mosaics and matched catalogs across GOODS-N, GOODS-S \citep{2004ApJ...600L..93G}, and COSMOS \citep{Capak_2007} provide high-resolution far-UV imaging over 44.7 arcmin$^2$ to typical depths of $\mathrm{FUV}\approx 28.7$~AB (3$\sigma$, $0.5"$ diameter aperture), with $\sim 1/3$ of the area deeper than 29.0, yielding 1068 far-UV detections primarily at $0.2<z<0.9$.

All images, segmentations, variance and time maps, and the matched catalog are released via MAST as a High-Level Science Product (HLSP) in formats aligned with GOODS/CANDELS/3D-HST/HLF, lowering the barrier to incorporate FUV diagnostics into multi-wavelength studies of galaxy evolution (\dataset[FUEL-SURVEY]{\doi{10.17909/2yqw-3g14}}).

These legacy products enable several community investigations in galaxy evolution:

\paragraph{Star-forming clumps: sizes and ages.}
Sub-arcsecond FUV imaging across three fields allows systematic identification and characterization of star-forming clumps (sizes and $L_{\rm FUV}$) and their evolution. With ancillary optical/IR and H$\alpha$ data, these images also enable clump-level estimates of star-formation rates, ages, and dust attenuation (e.g., \citealt{2019MNRAS.484.4897C, Martin_2023}). These data allow us to bridge the gap between local and $z>1$ already done with GALEX \citep{2005ApJ...619L...1M} and HST \citep{Teplitz2006,2007ApJ...668...62S}.

\paragraph{Resolved dust maps and attenuation curves.}
Combining FUV with deep near-UV/optical/near-IR imaging enables spatially resolved UV-slope and dust attenuation estimates. The mosaics support resolved analyses to test whether variations in attenuation curves or dust geometry drive observed UV-color gradients.

\paragraph{Timescales of ``bursty'' star formation.}
Joint use of a longer-timescale SFR tracer (FUV) with a shorter-timescale indicator (e.g., H$\alpha$ from ancillary spectroscopy or grism data) constrains burst amplitudes and duty cycles, especially in low-mass systems that dominate FUV-selected samples (e.g., \citealt{Weisz_2012, Emami_2019, 2015MNRAS.451..839D}).

\paragraph{Far-UV number counts and the UV EBL.}
These data enable measurements of the total ultraviolet extragalactic background light from galaxies, its evolution, and its effect on $\gamma$-ray attenuation, via differential and cumulative FUV number counts constructed from the three independent sightlines (e.g., \citealt{Xu_2005, Voyer_2011, Driver_2016}), and in the broader context of recent background/number-count analyses and systematics studies (e.g., \citealt{2018Sci...362.1031F,2024ApJ...972...95P,2022AJ....164..141W,10.1093/mnras/stag044}).

\paragraph{Lyman-continuum (LyC) escape.}
The F150LP bandpass probes rest-frame LyC for suitable redshift ranges (e.g., $z\sim1.2$--1.5) \citet{2007ApJ...668...62S}, \citet{Siana_2010}, \citet{2010ApJ...720..465B}, and \citet{Alavi_2020}. Uniform mosaics and astrometry enable stacked analyses of large, spectroscopically confirmed samples to constrain population-averaged LyC escape.

\section{Summary}
\label{sec:summary}

We present high-resolution \textit{HST}/ACS/SBC F150LP far-ultraviolet (FUV) images and photometric catalogs over three extragalactic sightlines: GOODS-S, GOODS-N, and COSMOS. We model and remove the spatially varying dark-current glow using masked stacks, exclude hot/bad pixels, register exposures to a GAIA-anchored frame, and produce mosaics.

The final SBC/F150LP mosaics cover a combined effective area of $44.7~\mathrm{arcmin}^2$ across the GOODS-N, GOODS-S, and COSMOS fields. Individual exposures are first subtracted of their uniform dark and dark glow. The corrected exposures are aligned to a common astrometric frame and drizzled on a single output grid (drizzled to a final pixel scale of $0.06''$ pixel$^{-1}$ for GOODS and $0.03''$ pixel$^{-1}$ for COSMOS), producing science mosaics together with the associated variance and exposure-time maps. The mosaics typically reach $\mathrm{FUV} \approx 28.7$~AB at $3\sigma$ in a $0.5^{\prime\prime}$-diameter aperture, enabling detection of faint rest-FUV sources in all three fields.

We perform photometry in both isophotal (defined by existing deep HST optical) and circular apertures. Astrometry is tied to \emph{Gaia} reference frames, and all FUV photometry includes Milky Way extinction corrections using a Fitzpatrick (1999) $R_V=3.1$ extinction law. Photometry is consistent with GALEX and previous SBC catalogs. The released catalogs report source positions, FUV magnitudes and uncertainties, a coverage flag, and also include cross-identifications to the 3D-HST and CANDELS catalogs. We distribute the field mosaics, variance maps, exposure-time maps, segmentation maps, and the photometric catalogs in FITS/CSV format, together with README files that describe the columns and flags. All data products are hosted at MAST as a High-Level Science Product (HLSP; \dataset[FUEL-SURVEY]{\doi{10.17909/2yqw-3g14}}).

The uniformly reduced mosaics and catalogs matched with other legacy data products enable a variety of science applications, including studies of rest-FUV emission from low- to moderate-redshift galaxies, sizes of star-forming clumps, and FUV number counts.

\begin{acknowledgments}
This research is based on observations made with the NASA/ESA Hubble Space Telescope obtained from the Space Telescope Science Institute, which is operated by the Association of Universities for Research in Astronomy, Inc., under NASA contract NAS 5--26555. These observations are associated with program(s) 9478, 10403, 10872, 11082, 11144, 11236, and 14123. 

Support for program \#17032 was provided by NASA through a grant from the Space Telescope Science Institute, which is operated by the Association of Universities for Research in Astronomy, Inc., under NASA contract NAS 5-03127.
\end{acknowledgments}

\bibliography{FUEL-SURVEY}{}
\bibliographystyle{aasjournal}

\end{document}